\documentclass[sn-mathphys,Numbered]{sn-jnl}

\usepackage{graphicx}%
\usepackage{multirow}%
\usepackage{amsmath,amssymb,amsfonts}%
\usepackage{amsthm}%
\usepackage{mathrsfs}%
\usepackage[title]{appendix}%
\usepackage{xcolor}%
\usepackage{textcomp}%
\usepackage{manyfoot}%
\usepackage{booktabs}%
\usepackage{algorithm}%
\usepackage{algorithmicx}%
\usepackage{algpseudocode}%
\usepackage{listings}%
\usepackage{siunitx}
\usepackage{lineno}
\usepackage{ulem}
\usepackage{url}

\begin{document}

\title[Evaluating the feasibility of short-integration scans based on the 2022 VGOS-R\&D program]{Evaluating the feasibility of short-integration scans based on the 2022 VGOS-R\&D program}

\author*[1]{\fnm{Matthias} \sur{Schartner}}\email{mschartner@ethz.ch}

\author[2]{\fnm{Bill} \sur{Petrachenko}}\email{wtpetra@gmail.com}
\equalcont{These authors contributed equally to this work.}

\author[3]{\fnm{Mike} \sur{Titus}}\email{matitus@mit.edu}
\equalcont{These authors contributed equally to this work.}

\author[4]{\fnm{Hana} \sur{Kr\'asn\'a}}\email{hana.krasna@geo.tuwien.ac.at}
\equalcont{These authors contributed equally to this work.}

\author[3]{\fnm{John} \sur{Barrett}}\email{barrettj@mit.edu}
\equalcont{These authors contributed equally to this work.}

\author[3]{\fnm{Dan} \sur{Hoak}}\email{dhoak@mit.edu}
\equalcont{These authors contributed equally to this work.}

\author[3]{\fnm{Dhiman} \sur{Mondal}}\email{dmondal@mit.edu}
\equalcont{These authors contributed equally to this work.}

\author[5]{\fnm{Minghui} \sur{Xu}}\email{minghui.xu@gfz-potsdam.de}
\equalcont{These authors contributed equally to this work.}

\author[1]{\fnm{Benedikt} \sur{Soja}}\email{soja@ethz.ch}
\equalcont{These authors contributed equally to this work.}

\affil*[1]{\orgdiv{Institute of Geodesy and Photogrammetry}, \orgname{ETH Zurich}, \orgaddress{\street{Robert-Gnehm-Weg 15}, \city{Zürich}, \postcode{8093}, \country{Switzerland}}}

\affil[2]{\orgname{Natural Resources Canada (retired)}}

\affil[3]{\orgname{MIT Haystack Observatory}}

\affil[4]{\orgname{TU Wien}}

\affil[5]{\orgname{DeutschesGeoForschungsZentrum (GFZ) Potsdam}}


\abstract{
In this work, we report on activities focusing on improving the observation strategy of the Very Long Baseline Interferometry (VLBI) Global Observing System (VGOS).
During six dedicated 24-hour Research and Development (R\&D) sessions conducted in 2022, the effectiveness of a signal-to-noise ratio (SNR)-based scheduling approach with observation times as short as 5--20 seconds was explored.
The sessions utilized a full \SI{8}{Gbps} observing mode and incorporated elements such as dedicated calibration scans, a VGOS frequency source-flux catalog, improved sky-coverage parameterization, and more. 

The number of scans scheduled per station increased by 2.34 times compared to operational VGOS-OPS sessions, resulting in a 2.58 times increase in the number of observations per station. 
Remarkably, the percentage of successful observations per baseline matched the fixed 30-second observation approach employed in VGOS-OPS, demonstrating the effectiveness of the SNR-based scheduling approach.

The impact on the geodetic results was examined based on statistical analysis, revealing a significant improvement when comparing the VGOS-R\&D program with VGOS-OPS. 
The formal errors in estimated station coordinates decreased by \SI{50}{\%}. 
The repeatability of baseline lengths improved by \SI{30}{\%}, demonstrating the enhanced precision of geodetic measurements. 
Furthermore, Earth orientation parameters exhibited substantial improvements, with a \SI{60}{\%} reduction in formal errors,  \SI{27}{\%} better agreement w.r.t. IVS-R1/R4, and \SI{13}{\%} better agreement w.r.t. IERS~EOP~20C04. 

Overall, these findings strongly indicate the superiority of the VGOS-R\&D program, positioning it as a role model for future operational VGOS observations.
}

\keywords{VLBI, VGOS, IVS}

\maketitle

\section{Introduction}

Very Long Baseline Interferometry (VLBI) is a cutting-edge technique in space geodesy. 
Through the synchronized observations of multiple radio telescopes strategically positioned worldwide, VLBI attains unrivaled accuracy in determining the rotation angle of the Earth about its axis and in monitoring minute variations of the orientation of the Earth's rotation vector in space. 
It also performs measurements that support the most precise operation of the Global Navigation Satellite System (GNSS) and makes important contributions to the establishment of the International Terrestrial Reference Frame \citep[ITRF;][]{Altamimi2023} while also defining the International Celestial Reference Frame \citep[ICRF; ][]{Charlot2020}. 
In this way, VLBI contributes to the precise measurement of Earth's tectonic plate movements, to the precise measurement of changes in global mean sea level, and to the study of other relevant geophysical processes. 

Over the decades, the geodetic VLBI technique has gone through several technical upgrades. 
The latest upgrade, originally called VLBI2010, is now referred to as the VLBI Global Observing System \citep[VGOS; ][]{Niell2006, Petrachenko2009, Niell2018}. 
The upgrade was motivated by a need to renew aging infrastructure and systems and more significantly to try to meet the demanding requirements of the Global Geodetic Observing System \citep[GGOS; ][]{Plag2009, Beutler2012}. 

During the VGOS design process, three sources of VLBI random errors were considered: delays through the neutral atmosphere, stochastic variations of the hydrogen maser reference oscillators, and measurement errors of the delay observable; and three sources of systematic errors were considered: the impact of source structure, uncalibrated drifts of the electronics, and gravitational and thermal deformations of the antenna structures. 
To study the random errors, extensive Monte-Carlo simulations were conducted \citep{Petrachenko2008, Pany2011}. 
As expected, these studies revealed that tropospheric turbulence is VLBI's primary random error source and furthermore that the impact of tropospheric turbulence can be significantly reduced by increasing the sampling rate of the troposphere. 
This requires rapid switching of the antennas between radio sources at different azimuth and elevation angles. 
VGOS specifies that each antenna should aim to observe on average one source every 30 seconds. 
To address this requirement, a new network of 20 or more smaller but faster slewing antennas, in the range of 12 to 13 meters in diameter, is in the process of being established. 
At the time of writing about 10 antennas are available operationally. 
To compensate for the reduced sensitivity of these smaller antennas, the recording rate has been increased to \SI{8}{Gbps}, distributed across four \SI{512}{MHz} bands A--D, with center frequencies approximately at \SI{3.2}{GHz}, \SI{5.5}{GHz}, \SI{6.6}{GHz}, and \SI{10.4}{GHz}. 
However, it is expected that in the future, the recording rate will be further increased to \SI{16}{Gbps}. 
It was estimated that the increase in recording rate would make it possible to detect more than 200 high-quality and spatially distributed radio sources using integration times no longer than 20 seconds, with stronger sources detected in 5 seconds or less.
Recent research by \citet{Anderson2018}, \citet{Xu2021a}, and \citet{Xu2021b} indicates that source structure effects also contribute significantly to the VGOS error budget. 
Work is underway to develop processes for imaging and modeling source morphology and for creating corrections for source structure.
This necessitates consideration during the generation of observing plans, i.e. the schedules of the sessions, as discussed in \citet{Schartner2023}. 

VGOS has already demonstrated impressive capabilities. 
Analyzing VGOS Intensive sessions, which are hour-long sessions designed for rapid determination of Earth's phase of rotation, \citet{Haas2021} revealed that the formal errors of VGOS were approximately three to four times better than those of traditional Intensives observed at S/X band, with an improved agreement with IERS~Bulletin~B \citep{Luzum2014}.
Most recently, \citet{Diamantidis2023} has analyzed the first three years of VGOS 24-hour sessions using a Kalman filter approach and revealed station position repeatabilities of around \SI{2}{mm} for the horizontal components and \SI{4}{mm} for the vertical component. 
However this study, as well as \citet{Glomsda2023} and \citet{ Nilsson2023} highlight there are still deficiencies of the polar motion and nutation estimates from VGOS, most probably related to the station distribution of the current VGOS network with all stations being located in the northern hemisphere and the resulting lack of long north-south baselines.
VGOS has further been successfully employed to determine vertical total electron content \citep[VTEC;][]{Motlaghzadeh2022} and tropospheric Zenith Wet Delays \citep[ZWD;][]{Haas2023}, exhibiting good agreement with Global Navigation Satellite Systems (GNSS) and other independent techniques such as water vapor radiometers in the case of ZWD.

Geodetic VLBI observations, including VGOS, are organized by the International VLBI Service for Geodesy and Astrometry \citep[IVS;][]{Nothnagel2017}. 
During 2022, six dedicated research and development (R\&D) sessions were scheduled specifically to investigate improvements in the VGOS technique.
These sessions were primarily aimed at increasing the source switching rate of VGOS observations. 
In current VGOS operations, hereinafter referred to as VGOS-OPS, source switching is achieved at most every 68 seconds, which is still less than half the specified rate of 30 seconds. 
Time allotted to a source observation includes: time required to slew the antenna to the source, time to carry out system checks and resets, idle time while the network arrives on source, 
time to do calibrations and finally the time on source, the integration time, when data is actually being acquired. 
In addition, at some stations, a time (equal to the data acquisition time) is required to flush a buffer. 
At these stations, the record rate is \SI{4}{Gbps}, which is half the acquire rate, hence the need for equal time to flush the buffer.

So the goal of the 2022 VGOS-R\&D sessions is to study each one of these time components and find ways to reduce them. 
Two strategies provided the greatest improvement. 
Firstly, VGOS-OPS uses a fixed data integration time of \SI{30}{s}, regardless of the strength of the radio source. 
In the R\&D sessions, a range of integration times is used depending on the strength of the source and the sensitivities of the antennas. 
In other words, integrations are shorter for stronger sources and vice versa.
This is referred to as SNR-driven scheduling.
In most sessions, the range of integration times was \SIrange{7}{20}{s}, while in one session it was reduced to \SIrange{5}{18}{s}. 
In general, the average integration time was about 11 seconds, which is nearly a third of the corresponding time for VGOS-OPS. 
Secondly, the record rate at the 'buffer-flush' stations could be doubled to \SI{8}{Gbps} by adding a second record module. 
This mode was used for all R\&D sessions. 
In addition, the time required for pre-scan calibration was reduced from \SI{4}{s} to \SI{2}{s} and \SI{4}{s} reserved for field system commands was completely eliminated.

In this work, we will report on the actions and investigations of the VGOS-R\&D sessions. 
Section~\ref{sec:sessions} describes the session designs, Session~\ref{sec:results} discusses the evaluation metrics and presents results, and Section~\ref{sec:conclusion} concludes the work. 

\section{Data} \label{sec:sessions}
The VGOS-R\&D sessions in 2022 were conducted at a bi-monthly frequency (i.e. one session every second month). 
Each session was individually designed and further discussed and approved by the VGOS Technical Committee (VTC). 
Due to a substantial backlog in VGOS correlation, results from previous sessions were not available before the subsequent sessions were observed. 
Hence, it was not possible to build on previous session results. 
Adjustments made between sessions could only be based on VTC discussions and log files obtained during observations. 

Unlike VGOS-OPS, the VGOS-R\&D schedules were generated using VieSched++ \citep{Schartner2019}. 
A major change from VGOS-OPS was the introduction of the SNR-based scheduling approach. 
The schedules were generated following \citet{Schartner2020a}.

Table~\ref{tab:vr22_base_stats} lists the start time and station network of each session.
Stations Hb and Nn were initially observed in tagalong mode as they were relatively new and their performance had not been fully validated. 
Tagalong mode refers to a scheduling technique, where the observing plan is first generated without the tagalong stations before adding them to the existing schedule. 
This way, losing the tagalong stations will not impact the schedule of the remaining stations. 
Similarly, tagalong mode was occasionally utilized for Oe and Ow due to potential storage limitations affecting the station's ability to observe. 
In the last session, station Mg operated in tagalong mode due to technical issues. 
As presented in Table~\ref{tab:vr22_base_stats}, all sessions, except VR2203, experienced station losses from the core network due to technical problems. 
Consequently, the generated schedule could not be fully executed as intended, leading to limitations in data interpretability to some extent. 

\begin{table}[htb]
    \caption{List of 2022 VGOS-R\&D sessions. 
    A full list of station names can be found on the IVS homepage\textsuperscript{1}. 
    Underlined stations were scheduled in tagalong mode. 
    Striked-through stations were included in the schedule but did not observe due to technical issues. 
    }
    \label{tab:vr22_base_stats}
    \centering
    \begin{tabular}{lll}
    \toprule
         Session & Start & Network \\
    \midrule
         VR2201 & 2022-01-20 18:00 & Gs \uline{Hb} K2 Mg Oe Ow Wf Ws \sout{Yj} \\
         VR2202 & 2022-03-17 18:00 & Gs K2 Mg Oe Wf Yj \uline{\sout{Hb}} \uline{\sout{Ow}} \sout{Ws} \\
         VR2203 & 2022-05-19 18:00 & Gs Is K2 Mg Ow Wf Yj \uline{\sout{Oe}} \\
         VR2204 & 2022-07-21 18:00 & Gs \uline{Hb} K2 Mg Oe \uline{Ow} Wf Ws Yj \sout{Is} \\
         VR2205 & 2022-09-15 18:00 & Hb Is K2 Mg \uline{Nn} Oe \uline{Ow} Wf Ws Yj \sout{Gs} \\
         VR2206 & 2022-11-09 18:00 & Hb K2 \uline{Mg} Nn Oe Wf Ws \sout{Gs} \uline{\sout{Ow}} \sout{Yj} \\         
    \bottomrule
    \end{tabular}
    
    \raggedleft
    \small\textsuperscript{1} https://cddis.nasa.gov/archive/vlbi/ivscontrol/ns-codes.txt
\end{table}

In the following, the most important changes in the session design are briefly summarized. 
The full scheduling setup is available in the operations notes accessible via the IVS data centers. 
Each subsequent session also includes the changes mentioned in previous sessions. 
Table~\ref{tab:design} provides a high-level overview of the session design.
\subsection{VR2201}
The first VGOS-R\&D session of 2022 was scheduled with a minimum scan time of seven seconds and a maximum scan time of 20 seconds. 
The target SNR was set to 10 for bands A--C and increased to 15 for band D. 
The band D increase was done to provide a margin for detection since band-D fluxes are typically weaker and band-D antenna sensitivities are typically lower. 
Due to a lack of source flux density information for the VGOS frequencies, the flux information was inter-/extrapolated from the S/X flux catalog using a source spectral index derived from the S/X fluxes.
\begin{align}
    S(\lambda_i) =& \frac{S_X}{\lambda_X^\alpha} \cdot \lambda^\alpha_i\\
    \alpha =& \frac{\log(S_X/S_S)}{\log(\lambda_X/\lambda_S)}
\end{align}
where $\lambda$ stands for the wavelength and $S$ stands for the source flux density. 
The information from the bands S and X ($S_S$, $S_X$, $\lambda_S$, $\lambda_X$) were used to derive the source flux density at the VGOS-frequency $\lambda_i$.
The indices represent the corresponding wavelength. 

The scheduling parameters were optimized using an evolutionary algorithm as outlined in \citet{Schartner2021b}.

In total, the initial source list covered 200 sources out of which 176 were observed during the session. 
25 of the 200 sources were scheduled with a fixed 20-second long observation time due to unreliable flux information and 17 of these 25 sources were observed in the final schedule. 
Almost all sources were observed a minimum of 10 times, well distributed over time, to cover different parallactic angles, ensure a good distribution of observations at the source UV plane required for imaging, and allow for estimating source coordinates. 


For correlation and fringe fitting purposes, 60-second-long calibrator scans were observed every two hours. 

\subsection{VR2202}
For VR2202, the initial source list was adjusted to 195 sources, out of which 141 were observed. 
The target SNR was reduced to 12 for band D. 
The main difference in scheduling was how the troposphere and in particular the sky-coverage optimization was performed. 
First, the VieSched++ software was extended to support individual sky-coverage parameters per station. 
These parameters include the influence distance between observations as well as the influence time between observations (for more information on these parameters see \citet[][Section 2.1.2]{Schartner2019}). 
Next, during the simulations, the ZWD and tropospheric gradients were estimated every 8 to 15 minutes, depending on the station. 
In contrast, during VR2201, as well as during scheduling simulations of S/X sessions, ZWD is typically estimated every 30 minutes and tropospheric gradients only every 180 minutes. 
Based on these changes, the Monte-Carlo-based scheduling optimization increased the importance of the sky-coverage optimization to \SI{60}{\%} instead of the \SI{3}{\%} used for VR2201.

To improve the inclusion of the only southern-hemisphere station Hb, sources with a declination $<$\ang{0} were allowed to be observed with two stations, instead of three stations. 

Finally, VR2202 observed a slightly different frequency setup compared to other VGOS sessions. 
While frequencies at bands A and D were only changed marginally by some tens of MHz in some channels, band B was offset by \SI{-768}{MHz} and band C was offset by \SI{+224}{MHz}. 
The reason for the change was to test a new frequency sequence that is expected to be more robust for fringe detection. 
Although fringe detection remains an important consideration for designing VGOS frequency sequences, constraints imposed by RFI need to be given priority. 
Hence, it was decided to defer further frequency sequence investigations until current and future RFI conditions are better understood over the VGOS network, including both currently operational stations and stations either undergoing commissioning or under construction. 

\subsection{VR2203}
For VR2203, the standard VGOS frequency setup was utilized again. 
This was done to guarantee successful observations since results from VR2202 were not yet available. 
However, there were two noteworthy adjustments to the session design of VR2203. 
First, the rate of calibration scans was doubled to one scan every hour based on a request by the correlator. 
Second, a new source flux density catalog was used for the first time. 
The new catalog was derived from previous VGOS-OPS and VGOS-R\&D sessions and provided fluxes directly for VGOS bands A, B, C, and D. 
This eliminated the need to interpolate from the S/X flux density catalog.
The new catalog included 176 sources. 
The initial source list was set to these sources, out of which 90 were observed. 

\subsection{VR2204}
In VR2204, the minimum observation time was reduced to 5 seconds while the maximum time was reduced to 18 seconds. 
Simultaneously, the target SNR level was increased to 12 for all bands. 
The source list was reduced to 129 well-performing sources out of which 85 were observed. 
Furthermore, the calibration scans were improved based on further feedback from the correlator and the VTC. 
The new calibrator scans allowed for subnetting of the network, to improve the likelihood that all stations are scheduled in each calibration block. 
Consequently, there were two calibrator scans (with the subnetted network) per hour in almost all cases. 

\subsection{VR2205}
In this session, the source list was again revised to include only the 100 best-performing sources, selected based on the percentage of successful observations and the root mean square error of the residuals. 
Furthermore, the SNR targets were individually adjusted per baseline and per frequency band based on the performance in the previous sessions. 
The minimum and maximum observing durations were returned to the initial 7 and 20 seconds respectively since results from VR2204 were not yet available to validate the 5-second long scans. 

\subsection{VR2206}
During the generation of VR2205 it became apparent that for a proper optimization of the schedule, more sources are required. 
Furthermore, based on VTC discussions, it was recommended that a better tie to the ICRF3 could be achieved if more ICRF3-defining sources were observed.
Thus, the final VR2206 session added an additional 41 ICRF3-defining sources to the initial source list. 
These sources were observed with a fixed 30-second long integration time instead of an SNR-based observation duration. 

\begin{table}[htb]
    \caption{Summary of most important session design decisions. }
    \label{tab:design}
    \centering
    \begin{tabular}{l c c c c c c}
    \toprule
          & VR2201 & VR2202 & VR2203 & VR2204 & VR2205 & VR2206 \\
    \midrule
        observation duration [s] & 7--20 &  7--20 &  7--20 &   5--18 &  7--20 & 7--20/30 \\
        calibrator scan spacing [hours] & 2 & 2 & 1 & 1 & 1 & 1 \\
        subnetted calibrator scans  &   &   &   & \checkmark & \checkmark & \checkmark \\
        improved sky-coverage param. &   &  \checkmark &  \checkmark & \checkmark & \checkmark & \checkmark \\
        VGOS source flux catalog &   &   &  \checkmark & \checkmark & \checkmark & \checkmark \\
        SNR-targets per freq. and baseline &   &   &   &  & \checkmark & \checkmark \\
        testing new ICRF3 defining sources &   &   &   &  &  & \checkmark \\
        more robust frequency sequence &   &  \checkmark &   &  &  & \\
    \bottomrule
    \end{tabular}
\end{table}


\section{Results and Discussion} \label{sec:results}

This section provides a concise summary of the scheduling statistics and analysis setup employed in the study. 
There are two data sources for obtaining the geodetic results. 
First, our own analysis, derived from group delays provided as databases via IVS Data Centers by using the Vienna VLBI and Satellite Software \citep[VieVS; ][]{Boehm2018}. 
Second, the official IVS analysis results that were obtained via the $\nu$-Solve package \citep{Bolotin2014}. 

Comparative analyses between VGOS-R\&D and VGOS-OPS sessions are performed, with the VGOS-OPS reference based on the sessions conducted in 2022 available at the time of writing this manuscript. 
Specifically, the VGOS-OPS dataset includes 39 sessions, ranging from VO2013 to VO2363 (2022-01-13 to 2022-12-29), excluding VO2286, VO2307, and VO2335 for which the databases have not been released at the time the study was conducted.

\subsection{Scheduling statistics}
\begin{figure}[htb]
    \centering
    \includegraphics[width=\textwidth]{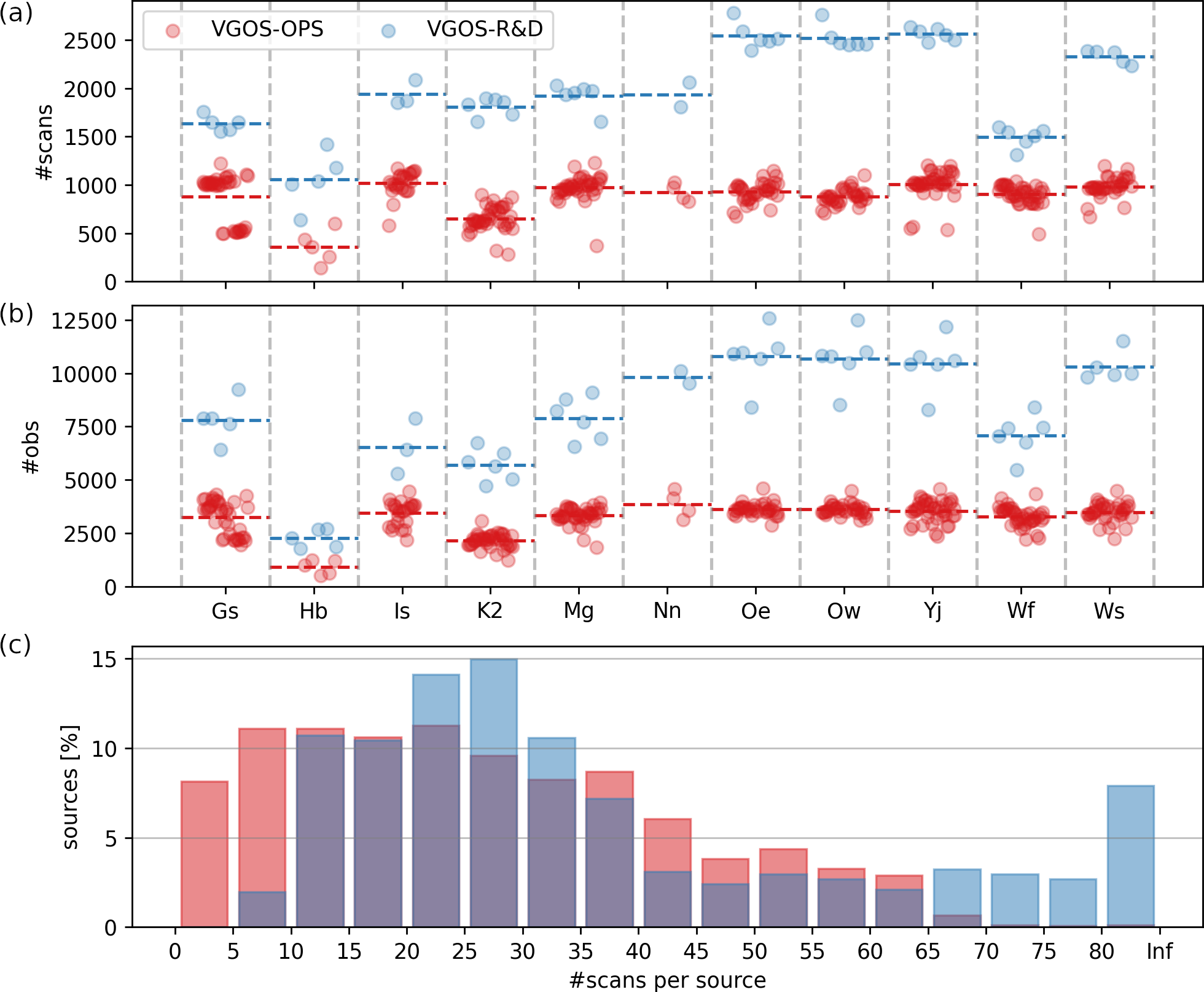}
    \caption{Scheduling statistics; (a) number of scans per station; (b) number of observations per station; (c) number of scans per source. 
    The markers in (a) and (b) represent individual sessions while the dashed lines depict the average. VGOS-R\&D sessions are depicted in blue while VGOS-OPS sessions are depicted in red. }
    \label{fig:stats}
\end{figure}

Figure~\ref{fig:stats} presents, for each station, the number of scans (a) and the number of observations (b) per session. 
The average number is indicated by the dashed line.
Considering VGOS-R\&D relative to VGOS-OPS, the number of scans per station, as well as the number of observations per station, is increased substantially. 
The increases are by a factor of 2.34 and 2.58 respectively. 

Figure~\ref{fig:stats} further depicts the number of scans per source. 
This metric is important for astrometry and imaging. 
It can be seen that the number of sources with less than ten scans is reduced significantly, improving the capability to estimate source positions during single-session analysis. 
This is discussed in more detail in \citet{Schartner2023}. 

\begin{figure}[htb]
    \centering
    \includegraphics[width=\textwidth]{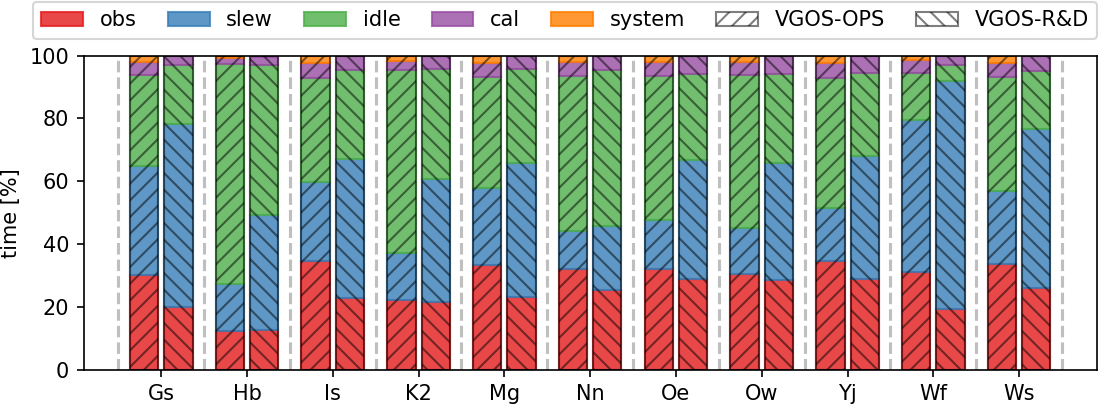}
    \caption{ Station by station time distribution of activities; obs: observation time, slew: slewing time, idle: station idling, cal: calibration time, system: time for field-system commands. The VGOS-OPS sessions are depicted in the left columns (with /-hash marks) while the VGOS-R\&D sessions are depicted in the right columns (with \textbackslash -hash marks). }
    \label{fig:time}
\end{figure}

Figure~\ref{fig:time} depicts, on a station-by-station basis, the distribution of time spent in different activities, e.g. observing (obs), slewing (slew), idling (idle), calibration (cal), and execution of field-system commands (system). 
The first noteworthy conclusion is that the VGOS-R\&D sessions have 19\% less observing time compared to VGOS-OPS, although 2.34 more scans were observed as discussed previously. 
This is due to the SNR-based scheduling algorithm. 
The reduction in observing time also implies less data transfer and fewer bits correlated, which is important because data transfer and correlation are the current operational bottlenecks of today's VGOS observations. 
Next, one can see that the slewing time is increased by a factor of 2.09. 
This indicates that the atmospheric sampling is improved since longer slewing times mean that more different azimuth and elevation angles are observed.
This achievement is especially noteworthy since within VGOS-OPS, the stations get 30 seconds of free slewing time during the buffer-flush time. 
Finally, one can see that there is still significant idle time left. 
However, idle time is always correlated with slewing time since the stations have to wait for the slowest station to finish slewing before starting observations. 
In the VGOS network, the slowest station is Wf which has almost no idle time. 

\subsection{Successful observations}
\begin{figure}
    \centering
    \includegraphics[width=.75\textwidth]{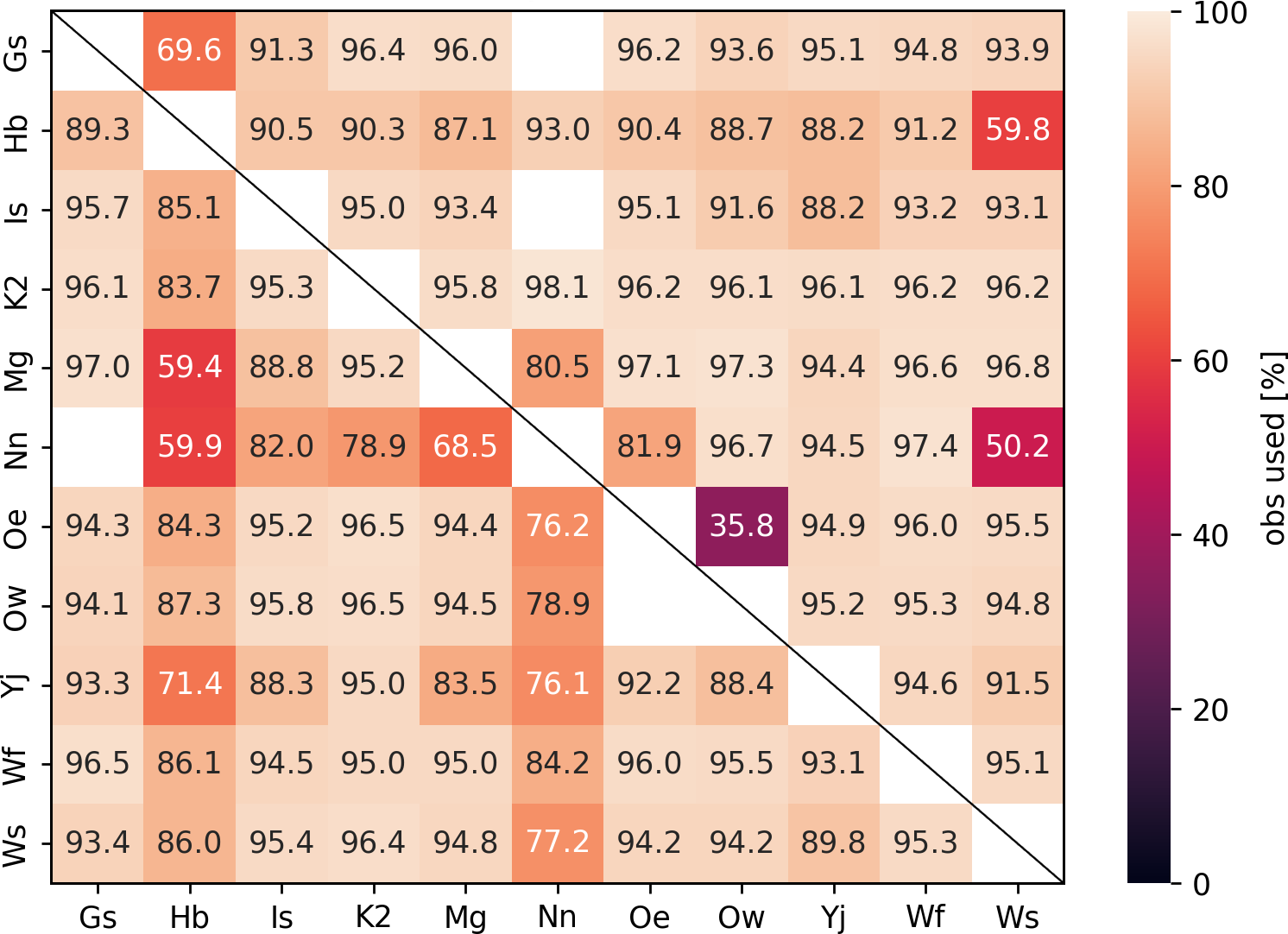}
    \caption{Percentage of successful observations used per baseline (0.75-quantile over all sessions). 
    Lower-left triangle: VGOS-R\&D. 
    Upper-right triangle: VGOS-OPS. }
    \label{fig:obs_success}
\end{figure}

The main research question of the VGOS-R\&D sessions was the feasibility of the short, SNR-based observation durations. 
Figure~\ref{fig:obs_success} gives an overview of the percentage of successful observations per baseline and compares it with the 2022 VGOS-OPS sessions.
In this context, we define a successful observation as an observation used in the geodetic analysis. 
The values were extracted from the official IVS analysis reports. 
The lower left triangle lists the values for VGOS-R\&D while the upper right triangle depicts the values for VGOS-OPS. 
The values listed are the 0.75-quantiles over all sessions. 
Most importantly, it can be seen that for almost all baselines, the difference between VGOS-R\&D and VGOS-OPS is insignificant (the median of VGOS-OPS is \SI{95}{\%} while the median of VGOS-R\&D is \SI{93}{\%}). 
This proves that the SNR-based observation times are technically possible and that the percentage of successful observations is comparable with VGOS-OPS. 
When keeping in mind that VGOS-R\&D scheduled a factor of 2.58 times more observations, one can conclude that the total yield of usable observations for analysis is significantly improved. 

Some outlier values exist that can easily be explained since they are either related to the fact that the VGOS-R\&D sessions included new and, at the time of the sessions, still not fully validated stations Hb and Nn, or to the local baseline Oe/Ow. 
It is also to note that a certain amount of observations might have failed not because of too low SNR but rather because of other technical or environmental reasons which is not reflected in this analysis.
This is also the reason why a robust metric (the 0.75-quantile) is depicted in Figure~\ref{fig:obs_success} instead of the mean value to mitigate this effect. 

\subsection{Geodetic results}

In the following, the geodetic results from VGOS-R\&D are compared with the results from VGOS-OPS. 
It is important to interpret the comparisons between VGOS-R\&D and VGOS-OPS carefully due to variations in the network geometries, technical upgrades at the stations, and the substantial disparity in the number of sessions between the two datasets.
Additionally, the frequent occurrence of station dropouts further adds complexity to the comparisons. 
However, as demonstrated in this section, clear trends can be identified. 

In our own analysis, the a priori group delays were modeled as described in \citet{Krasna2023a}. 
Station position time series were obtained from solutions where the source coordinates were fixed to the ICRF3, while the station coordinates, as well as datum definition, were based on a priori information from the International Terrestrial Reference Frame 2020. 
Tropospheric parameters, including ZWD and gradients, were estimated every 30 minutes with relative constraints of \SI{0.5}{mm} between the piece-wise linear offsets (PWLO). 
Earth orientation parameters (EOP) were estimated as PWLO at 00:00~UTC. 
Polar motion and UT1-UTC were constrained relatively with \SI{10}{mas}, while celestial pole offsets were tightly constrained with \SI{0.1}{\micro as} over the session, which effectively delivers one offset at the mid-epoch of the session, approximately around 06:00 UTC.

\begin{figure}[htb]
    \centering
    \includegraphics[width=\textwidth]{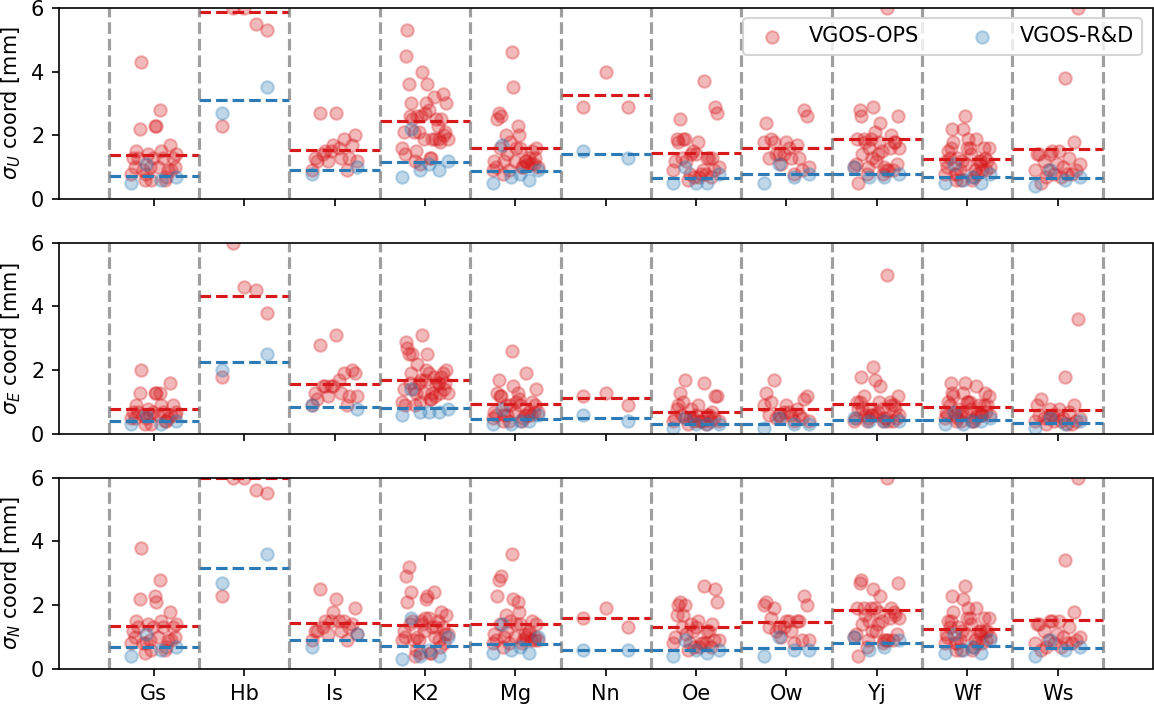}
    \caption{ Mean formal errors ($\sigma$) of station coordinates in up (top), east (mid), and north (bottom). The markers represent individual sessions while the dashed lines depict the average. 
    VGOS-R\&D results are depicted in blue while VGOS-OPS results are depicted in red. 
    Values larger than \SI{6}{mm} are displayed at \SI{6}{mm}.}
    \label{fig:coord}
\end{figure}

Figure~\ref{fig:coord} depicts the formal errors $\sigma$ of the estimated station coordinates based on the VieVS analysis. 
The average reduction of the mean formal errors per station is \SI{51}{\%}, \SI{52}{\%}, and \SI{50}{\%} in the up-, east-, and north-component, respectively. 
Based on the results obtained from the IVS analysis reports, the improvements are \SI{43}{\%}, \SI{35}{\%}, and \SI{38}{\%} for the up-, east-, and north-component, respectively. 
Thus, both analysis approaches confirm a significant reduction of formal errors. 

Next, the baseline-length-repeatability (blr) is investigated based on the VieVS analysis results. 
\begin{figure}[htb]
    \centering
    \includegraphics[width=\textwidth]{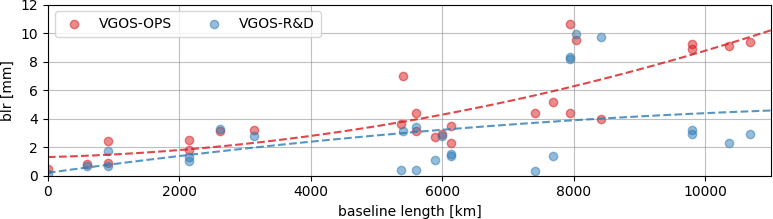}
    \caption{ Baseline length repeatability (blr) based on 26 baselines observed at least three times during VGOS-R\&D. 
    VGOS-R\&D results are depicted in blue while VGOS-OPS results are depicted in red. 
    The dashed lines represent a polynomial fit of degree two. 
    }
    \label{fig:blrep}
\end{figure}
In total, the investigation covers 26 baselines that have been analyzed in at least three VGOS-R\&D sessions. 
Thus, in particular, it does not include results from the new stations Nn and Hb, station Is, as well as Gs-Ws and Yj-Ws. 

The reduction of the blr is on average \SI{1.7}{mm} or \SI{30}{\%}.
The integrated polynomial fit of order two between \SIrange{0}{11000}{km} results in the same improvement of \SI{1.7}{mm} while the improvement at \SI{10000}{km} is \SI{50}{\%}, from \SI{8.8}{mm} to \SI{4.4}{mm}. 
Significant improvement could be seen for baselines including station K2 and European stations (around \SI{10000}{km} long) and K2 and the other American stations (around \SI{7500}{km} long). 
However, there is an increase in the repeatability for baselines including Mg and European stations (around \SI{8000}{km} long). 
The reason for this increase is a significant difference of several centimeters in the Mg y-coordinate between June 23rd and July 28th compared to the rest of 2022. 
The cause of this difference is unclear and currently under further investigation. 
Since VR2204 falls into this period and the VR sample size is small, it highly affects the VGOS-R\&D blr for the baselines involving Mg. 

\begin{figure}[htb]
    \centering
    \includegraphics[width=\textwidth]{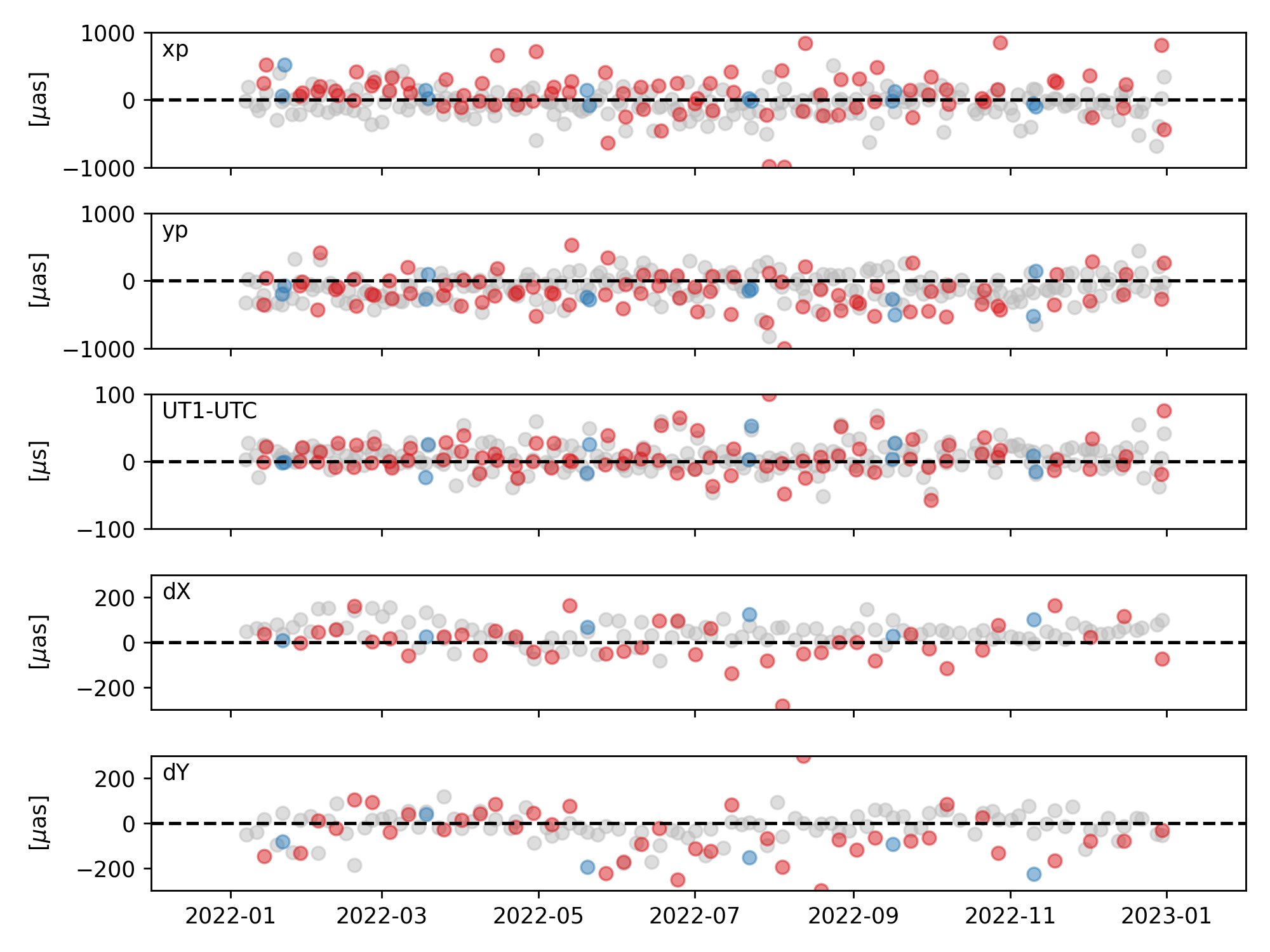}
    \caption{ Estimated EOP w.r.t. IERS~EOP~20C04 solution.
    From top to bottom: polar motion in x- (xp) and in y-direction (yp), UT1-UTC, celestial pole offsets in x- (dX), and in y-direction (dY).
    The black dashed line represents perfect agreement with the IERS solution. 
    VGOS-R\&D results are depicted in blue while VGOS-OPS results are displayed in red. 
    The IVS-R1/R4 results are depicted in gray. 
    }
    \label{fig:eop}
\end{figure}

Finally, the estimated EOP are compared. 
The EOP presented in Figure~\ref{fig:eop} are consistent with global solution VIE2022 \citep{Krasna2023} which produced an updated TRF and CRF including the most recent sessions. 
The adjustment involved 7308 24h IVS S/X and VGOS sessions from 1979.6 until 2023.0. 
By comparing the reported formal errors of the EOP estimates, an average improvement of \SI{49}{\%} could be identified, \SI{49}{\%} for polar motion in x-direction (xp), \SI{51}{\%} for polar motion in y-direction (yp), \SI{51}{\%} for UT1-UTC, \SI{47}{\%} for celestial pole offsets in x-direction (dX) and \SI{46}{\%} for celestial pole offsets in y-direction (dY). 
The improvement in terms of the formal errors of the EOP is in good agreement with the improvement in terms of the formal errors of the station coordinates (\SIrange{50}{52}{\%}) discussed earlier. 

Next, the EOP estimates from VGOS-R\&D and VGOS-OPS are compared with the estimates provided in the IERS~EOP~20C04 solution as well as with estimates obtained from IVS-R1/R4 sessions. 
The average improvement of the EOP estimates in terms of root mean squared is \SI{13}{\%} when considering IERS~EOP~20C04 as the reference and \SI{27}{\%} when considering IVS-R1/R4 as the reference.
Figure~\ref{fig:eop} depicts the agreement of the estimated EOP with the IERS~EOP~20C04 solution and the IVS-R1/R4 solution. 
The improvement of VGOS-R\&D compared to VGOS-OPS when considering IERS~EOP~20C04 as the reference in terms of root mean squared is \SI{49}{\%}, \SI{8}{\%}, \SI{16}{\%}, \SI{14}{\%} , and \SI{-21}{\%} for xp, yp, UT1-UTC, dX, and dY, respectively.

Similarly, the improvement when considering IVS-R1/R4 as the reference is \SI{36}{\%}, \SI{15}{\%}, \SI{43}{\%}, \SI{34}{\%}, and \SI{8}{\%} for xp, yp, UT1-UTC, dX, and dY, respectively. 

Thus, in all investigations, except for dY when considering IERS~EOP~20C04 as a reference, the VGOS-R\&D solutions show improvements over the VGOS-OPS solution. 
However, it is to be noted that the current VGOS network is not yet optimal for estimating EOP since almost all stations except for Hb are located in the northern hemisphere. 
Consequently, there are no long north-south baselines that are especially important for estimating polar motion and the celestial pole offsets. 
Furthermore, there are only a few observations of radio sources in the southern hemisphere, or even close to the equator, which also compromises the sensitivity to celestial pole offsets. 

\section{Conclusions} \label{sec:conclusion}
During 2022, the IVS provided resources for six 24-hour VGOS-R\&D sessions. 
These sessions were conducted with the objective of enhancing the VGOS observation strategy, specifically by evaluating an SNR-based scheduling approach with shorter observation times.
The VGOS-R\&D sessions were executed using the full \SI{8}{Gbps} observing mode, eliminating the need for buffer flush delays. 
Subsequent adjustments to the observation strategy were explored encompassing dedicated calibration scans, a source-flux catalog tailored for VGOS frequencies, improved sky-coverage parameterization, and revisions to the field-system commands.

The outcome of these efforts demonstrated a notable increase in the number of scans scheduled per station, reaching 2.34 times more than the operational VGOS-OPS sessions. 
Consequently, the number of observations per station increased by a factor of 2.58. 
Importantly, we have confirmed that the percentage of successful observations per baseline is on par with the established 30-second fixed observation approach utilized in VGOS-OPS, assuming that no technical errors occur, thus validating the efficacy of the SNR-based scheduling approach.
Previous investigations have already shown that VGOS-R\&D sessions yield improved tropospheric sampling \citep{Haas2023}, and in this study, we have examined their impact on geodetic results.
Despite the relatively small sample size of only six sessions, the evidence strongly indicates the superiority of the VGOS-R\&D program over VGOS-OPS. 
Notably, the formal errors in the estimated station coordinates were reduced by \SI{50}{\%}, while the repeatability of baseline lengths improved by \SI{30}{\%}.
The EOP formal errors were reduced by {49}{\%}, while improving the agreement between VGOS-R\&D and IVS-R1/R4 by \SI{27}{\%}. 
The agreement with respect to IERS~EOP~20C04 exhibited improvements of \SI{13}{\%}. 
Consequently, we conclude that the VGOS-R\&D program has been successful and can serve as a role model for future VGOS observations.

Nevertheless, certain challenges have surfaced, necessitating further sessions to solidify our claim of the superiority of VGOS-R\&D. 
Technical problems led to station dropouts in all sessions, except for VR2203, resulting in deviations from the originally intended schedule. 
Furthermore, the limited sample size and extensive number of adjustments implemented make it challenging to identify the specific contribution of each modification.
Thus, evaluation of the VGOS-R\&D program as a whole is the only feasible approach. 
In particular, although it was demonstrated that observing an altered frequency setup is technically possible, no conclusion regarding its performance can be drawn since only one session has been observed. 
Furthermore, it became evident that calibrator scans every hour that cover all stations are beneficial based on reports from the correlator staff. 

From an operational perspective, it is worth noting that most of the changes implemented in VGOS-R\&D can be readily transferred to VGOS-OPS. 
The only exception pertains to the individual SNR targets per frequency and baseline introduced in VR2205, which necessitate substantial manual work during the scheduling setup. 
However, these adjustments were deemed excessive, as practical observations generally exhibit significantly higher SNR levels than the targeted SNR due to the selection of the least sensitive baseline and frequency pair to set each scan's total duration. 
Thus, using these individual SNR targets is abandoned in the 2023 VGOS-R\&D program.
Lastly, the source flux catalog for VGOS frequencies would require regular updates akin to its S/X counterpart for operational usage.

\section*{Statements and Declarations}
\subsection*{Availability of data and materials}
The IVS-related datasets generated and/or analyzed during the current study are available in the IVS data centers and can be accessed via the corresponding session listed in  \url{https://ivscc.gsfc.nasa.gov/sessions/2022/}. 
The EOP dataset VIE2022 analyzed in this paper is available at \url{https://doi.org/10.48436/0gmbv-arv60}.

\subsection*{Competing interests}
The authors declare that they have no competing interests.

\subsection*{Funding}
The work by MIT Haystack Observatory was supported under NASA contract Awarding Agency.
\ \\
Open access funding is provided by the Swiss Federal Institute of Technology Zurich.

\subsection*{Authors' contributions}
MS and BP designed the concepts of the VGOS-R\&D sessions with support from MX. 
MS generated the observing plans.
MT correlated the sessions. 
MT, DH, and JB performed the fringe-fitting and post-correlation-processing.
DM performed data quality checks with support from DH, MT, and JB. 
HK performed the geodetic analysis using VieVS. 
BP developed monitoring software to analyze the performance of stations and the brightness of sources. 
MS wrote the majority of the manuscript and summarized, compared, and visualized the results. 
All authors read and contributed to the manuscript. 

\subsection*{Acknowledgments}
This research has made use of VGOS data files provided by the International VLBI Service for Geodesy and Astrometry (IVS) data archives, dated July 2023.
\ \\
We acknowledge the assistance of AI language model ChatGPT in providing suggestions on improved text formulations during the preparation of this paper.

\bibliography{sn-bibliography}


\begin{thebibliography}{28}
\ifx \bisbn   \undefined \def \bisbn  #1{ISBN #1}\fi
\ifx \binits  \undefined \def \binits#1{#1}\fi
\ifx \bauthor  \undefined \def \bauthor#1{#1}\fi
\ifx \batitle  \undefined \def \batitle#1{#1}\fi
\ifx \bjtitle  \undefined \def \bjtitle#1{#1}\fi
\ifx \bvolume  \undefined \def \bvolume#1{\textbf{#1}}\fi
\ifx \byear  \undefined \def \byear#1{#1}\fi
\ifx \bissue  \undefined \def \bissue#1{#1}\fi
\ifx \bfpage  \undefined \def \bfpage#1{#1}\fi
\ifx \blpage  \undefined \def \blpage #1{#1}\fi
\ifx \burl  \undefined \def \burl#1{\textsf{#1}}\fi
\ifx \doiurl  \undefined \def \doiurl#1{\url{https://doi.org/#1}}\fi
\ifx \betal  \undefined \def \betal{\textit{et al.}}\fi
\ifx \binstitute  \undefined \def \binstitute#1{#1}\fi
\ifx \binstitutionaled  \undefined \def \binstitutionaled#1{#1}\fi
\ifx \bctitle  \undefined \def \bctitle#1{#1}\fi
\ifx \beditor  \undefined \def \beditor#1{#1}\fi
\ifx \bpublisher  \undefined \def \bpublisher#1{#1}\fi
\ifx \bbtitle  \undefined \def \bbtitle#1{#1}\fi
\ifx \bedition  \undefined \def \bedition#1{#1}\fi
\ifx \bseriesno  \undefined \def \bseriesno#1{#1}\fi
\ifx \blocation  \undefined \def \blocation#1{#1}\fi
\ifx \bsertitle  \undefined \def \bsertitle#1{#1}\fi
\ifx \bsnm \undefined \def \bsnm#1{#1}\fi
\ifx \bsuffix \undefined \def \bsuffix#1{#1}\fi
\ifx \bparticle \undefined \def \bparticle#1{#1}\fi
\ifx \barticle \undefined \def \barticle#1{#1}\fi
\bibcommenthead
\ifx \bconfdate \undefined \def \bconfdate #1{#1}\fi
\ifx \botherref \undefined \def \botherref #1{#1}\fi
\ifx \url \undefined \def \url#1{\textsf{#1}}\fi
\ifx \bchapter \undefined \def \bchapter#1{#1}\fi
\ifx \bbook \undefined \def \bbook#1{#1}\fi
\ifx \bcomment \undefined \def \bcomment#1{#1}\fi
\ifx \oauthor \undefined \def \oauthor#1{#1}\fi
\ifx \citeauthoryear \undefined \def \citeauthoryear#1{#1}\fi
\ifx \endbibitem  \undefined \def \endbibitem {}\fi
\ifx \bconflocation  \undefined \def \bconflocation#1{#1}\fi
\ifx \arxivurl  \undefined \def \arxivurl#1{\textsf{#1}}\fi
\csname PreBibitemsHook\endcsname

\bibitem[\protect\citeauthoryear{Altamimi et~al.}{2023}]{Altamimi2023}
\begin{botherref}
\oauthor{\bsnm{Altamimi}, \binits{Z.}},
\oauthor{\bsnm{Rebischung}, \binits{P.}},
\oauthor{\bsnm{Collilieux}, \binits{X.}},
\oauthor{\bsnm{M{\'{e}}tivier}, \binits{L.}},
\oauthor{\bsnm{Chanard}, \binits{K.}}:
{ITRF}2020: an augmented reference frame refining the modeling of nonlinear
  station motions.
Journal of Geodesy
\textbf{97}(5)
(2023)
\doiurl{10.1007/s00190-023-01738-w}
\end{botherref}
\endbibitem

\bibitem[\protect\citeauthoryear{Charlot et~al.}{2020}]{Charlot2020}
\begin{barticle}
\bauthor{\bsnm{Charlot}, \binits{P.}},
\bauthor{\bsnm{Jacobs}, \binits{C.S.}},
\bauthor{\bsnm{Gordon}, \binits{D.}},
\bauthor{\bsnm{Lambert}, \binits{S.}},
\bauthor{\bsnm{Witt}, \binits{A.}},
\bauthor{\bsnm{B\"{o}hm}, \binits{J.}},
\bauthor{\bsnm{Fey}, \binits{A.L.}},
\bauthor{\bsnm{Heinkelmann}, \binits{R.}},
\bauthor{\bsnm{Skurikhina}, \binits{E.}},
\bauthor{\bsnm{Titov}, \binits{O.}},
\bauthor{\bsnm{Arias}, \binits{E.F.}},
\bauthor{\bsnm{Bolotin}, \binits{S.}},
\bauthor{\bsnm{Bourda}, \binits{G.}},
\bauthor{\bsnm{Ma}, \binits{C.}},
\bauthor{\bsnm{Malkin}, \binits{Z.}},
\bauthor{\bsnm{Nothnagel}, \binits{A.}},
\bauthor{\bsnm{Mayer}, \binits{D.}},
\bauthor{\bsnm{MacMillan}, \binits{D.S.}},
\bauthor{\bsnm{Nilsson}, \binits{T.}},
\bauthor{\bsnm{Gaume}, \binits{R.}}:
\batitle{{The third realization of the International Celestial Reference Frame
  by very long baseline interferometry}}.
\bjtitle{Astronomy {\&} Astrophysics}
(\byear{2020})
\doiurl{10.1051/0004-6361/202038368}
\end{barticle}
\endbibitem

\bibitem[\protect\citeauthoryear{Niell et~al.}{2005}]{Niell2006}
\begin{botherref}
\oauthor{\bsnm{Niell}, \binits{A.E.}},
\oauthor{\bsnm{Whitney}, \binits{A.R.}},
\oauthor{\bsnm{Petrachenko}, \binits{B.}},
\oauthor{\bsnm{Schl{\"u}ter}, \binits{W.}},
\oauthor{\bsnm{Vandenberg}, \binits{N.R.}},
\oauthor{\bsnm{Hase}, \binits{H.}},
\oauthor{\bsnm{Koyama}, \binits{Y.}},
\oauthor{\bsnm{Schuh}, \binits{H.}},
\oauthor{\bsnm{Tuccari}, \binits{G.G.}}:
{VLBI2010}: current and future requirements for geodetic {VLBI} systems. Tech.
  Rep
(2005)
\end{botherref}
\endbibitem

\bibitem[\protect\citeauthoryear{Petrachenko et~al.}{2009}]{Petrachenko2009}
\begin{botherref}
\oauthor{\bsnm{Petrachenko}, \binits{B.}},
\oauthor{\bsnm{Niell}, \binits{A.}},
\oauthor{\bsnm{Behrend}, \binits{D.}},
\oauthor{\bsnm{Corey}, \binits{B.}},
\oauthor{\bsnm{B\"ohm}, \binits{J.}},
\oauthor{\bsnm{Charlot}, \binits{P.}},
\oauthor{\bsnm{Collioud}, \binits{A.}},
\oauthor{\bsnm{Gipson}, \binits{J.}},
\oauthor{\bsnm{Haas}, \binits{R.}},
\oauthor{\bsnm{Hobiger}, \binits{T.}},
\oauthor{\bsnm{Koyama}, \binits{Y.}},
\oauthor{\bsnm{MacMillan}, \binits{D.}},
\oauthor{\bsnm{Malkin}, \binits{Z.}},
\oauthor{\bsnm{Nilsson}, \binits{T.}},
\oauthor{\bsnm{Pany}, \binits{A.}},
\oauthor{\bsnm{Tuccari}, \binits{G.}},
\oauthor{\bsnm{Whitney}, \binits{A.}},
\oauthor{\bsnm{Wresnik}, \binits{J.}}:
{Design Aspects of the {VLBI}2010 System. Progress Report of the IVS {VLBI}2010
  Committee, June 2009.}
{NASA/TM}-2009-214180, 2009, 62 pages
(2009).
\url{https://hal.archives-ouvertes.fr/hal-00582342}
\end{botherref}
\endbibitem

\bibitem[\protect\citeauthoryear{Niell et~al.}{2018}]{Niell2018}
\begin{botherref}
\oauthor{\bsnm{Niell}, \binits{A.}},
\oauthor{\bsnm{Barrett}, \binits{J.}},
\oauthor{\bsnm{Burns}, \binits{A.}},
\oauthor{\bsnm{Cappallo}, \binits{R.}},
\oauthor{\bsnm{Corey}, \binits{B.}},
\oauthor{\bsnm{Derome}, \binits{M.}},
\oauthor{\bsnm{Eckert}, \binits{C.}},
\oauthor{\bsnm{Elosegui}, \binits{P.}},
\oauthor{\bsnm{McWhirter}, \binits{R.}},
\oauthor{\bsnm{Poirier}, \binits{M.}},
\oauthor{\bsnm{Rajagopalan}, \binits{G.}},
\oauthor{\bsnm{Rogers}, \binits{A.}},
\oauthor{\bsnm{Ruszczyk}, \binits{C.}},
\oauthor{\bsnm{SooHoo}, \binits{J.}},
\oauthor{\bsnm{Titus}, \binits{M.}},
\oauthor{\bsnm{Whitney}, \binits{A.}},
\oauthor{\bsnm{Behrend}, \binits{D.}},
\oauthor{\bsnm{Bolotin}, \binits{S.}},
\oauthor{\bsnm{Gipson}, \binits{J.}},
\oauthor{\bsnm{Petrachenko}, \binits{B.}}:
{Demonstration of a Broadband Very Long Baseline Interferometer System: A New
  Instrument for High-Precision Space Geodesy}.
Radio Science
\textbf{53}
(2018)
\doiurl{10.1029/2018RS006617}
\end{botherref}
\endbibitem

\bibitem[\protect\citeauthoryear{Plag and Pearlman}{2009}]{Plag2009}
\begin{bbook}
\bauthor{\bsnm{Plag}, \binits{H.-P.}},
\bauthor{\bsnm{Pearlman}, \binits{M.}}:
\bbtitle{Global Geodetic Observing System: Meeting the Requirements of a Global
  Society on a Changing Planet in 2020},
pp. \bfpage{1}--\blpage{332}.
\bpublisher{Springer}, \blocation{???}
(\byear{2009}).
\doiurl{10.1007/978-3-642-02687-4}
\end{bbook}
\endbibitem

\bibitem[\protect\citeauthoryear{Beutler and Rummel}{2012}]{Beutler2012}
\begin{bchapter}
\bauthor{\bsnm{Beutler}, \binits{G.}},
\bauthor{\bsnm{Rummel}, \binits{R.}}:
\bctitle{{Scientific Rationale and Development of the Global Geodetic Observing
  System}}.
In: \beditor{\bsnm{Kenyon}, \binits{S.}},
\beditor{\bsnm{Pacino}, \binits{M.C.}},
\beditor{\bsnm{Marti}, \binits{U.}} (eds.)
\bbtitle{Geodesy for Planet Earth},
pp. \bfpage{987}--\blpage{993}.
\bpublisher{Springer},
\blocation{Berlin, Heidelberg}
(\byear{2012})
\end{bchapter}
\endbibitem

\bibitem[\protect\citeauthoryear{Petrachenko et~al.}{2008}]{Petrachenko2008}
\begin{bchapter}
\bauthor{\bsnm{Petrachenko}, \binits{B.}},
\bauthor{\bsnm{Boehm}, \binits{J.}},
\bauthor{\bsnm{Macmillan}, \binits{D.}},
\bauthor{\bsnm{Niell}, \binits{A.}},
\bauthor{\bsnm{Pany}, \binits{A.}},
\bauthor{\bsnm{Searle}, \binits{A.}},
\bauthor{\bsnm{Wresnik}, \binits{J.}}:
\bctitle{Vlbi2010 antenna slew rate study}.
In: \beditor{\bsnm{Finkelstein}, \binits{A.}},
\beditor{\bsnm{{Behrend}}, \binits{D.}} (eds.)
\bbtitle{International {VLBI} Service for Geodesy and Astrometry 2008 General
  Meeting Proceedings},
pp. \bfpage{410}--\blpage{415}
(\byear{2008})
\end{bchapter}
\endbibitem

\bibitem[\protect\citeauthoryear{Pany et~al.}{2011}]{Pany2011}
\begin{barticle}
\bauthor{\bsnm{Pany}, \binits{A.}},
\bauthor{\bsnm{B{\"o}hm}, \binits{J.}},
\bauthor{\bsnm{MacMillan}, \binits{D.}},
\bauthor{\bsnm{Schuh}, \binits{H.}},
\bauthor{\bsnm{Nilsson}, \binits{T.}},
\bauthor{\bsnm{Wresnik}, \binits{J.}}:
\batitle{{Monte Carlo simulations of the impact of troposphere, clock and
  measurement errors on the repeatability of {VLBI} positions}}.
\bjtitle{Journal of Geodesy}
\bvolume{85}(\bissue{1}),
\bfpage{39}--\blpage{50}
(\byear{2011})
\doiurl{10.1007/s00190-010-0415-1}
\end{barticle}
\endbibitem

\bibitem[\protect\citeauthoryear{Anderson and Xu}{2018}]{Anderson2018}
\begin{barticle}
\bauthor{\bsnm{Anderson}, \binits{J.M.}},
\bauthor{\bsnm{Xu}, \binits{M.H.}}:
\batitle{{Source Structure and Measurement Noise Are as Important as All Other
  Residual Sources in Geodetic VLBI Combined}}.
\bjtitle{Journal of Geophysical Research: Solid Earth}
\bvolume{123}(\bissue{11}),
\bfpage{10162}--\blpage{10190}
(\byear{2018})
\doiurl{10.1029/2018JB015550}
{\href{https://arxiv.org/abs/https://agupubs.onlinelibrary.wiley.com/doi/pdf/10.1029/2018JB015550}{{https://agupubs.onlinelibrary.wiley.com/doi/pdf/10.1029/2018JB015550}}}
\end{barticle}
\endbibitem

\bibitem[\protect\citeauthoryear{Xu et~al.}{2021a}]{Xu2021a}
\begin{botherref}
\oauthor{\bsnm{Xu}, \binits{M.H.}},
\oauthor{\bsnm{Anderson}, \binits{J.M.}},
\oauthor{\bsnm{Heinkelmann}, \binits{R.}},
\oauthor{\bsnm{Lunz}, \binits{S.}},
\oauthor{\bsnm{Schuh}, \binits{H.}},
\oauthor{\bsnm{Wang}, \binits{G.}}:
Observable quality assessment of broadband very long baseline interferometry
  system.
Journal of Geodesy
\textbf{95}(5)
(2021)
\doiurl{10.1007/s00190-021-01496-7}
\end{botherref}
\endbibitem

\bibitem[\protect\citeauthoryear{Xu et~al.}{2021b}]{Xu2021b}
\begin{botherref}
\oauthor{\bsnm{Xu}, \binits{M.H.}},
\oauthor{\bsnm{Savolainen}, \binits{T.}},
\oauthor{\bsnm{Zubko}, \binits{N.}},
\oauthor{\bsnm{Poutanen}, \binits{M.}},
\oauthor{\bsnm{Lunz}, \binits{S.}},
\oauthor{\bsnm{Schuh}, \binits{H.}},
\oauthor{\bsnm{Wang}, \binits{G.L.}}:
{Imaging VGOS Observations and Investigating Source Structure Effects}.
Journal of Geophysical Research: Solid Earth
\textbf{126}(4)
(2021)
\doiurl{10.1029/2020jb021238}
\end{botherref}
\endbibitem

\bibitem[\protect\citeauthoryear{Schartner et~al.}{2023}]{Schartner2023}
\begin{botherref}
\oauthor{\bsnm{Schartner}, \binits{M.}},
\oauthor{\bsnm{Collioud}, \binits{A.}},
\oauthor{\bsnm{Charlot}, \binits{P.}},
\oauthor{\bsnm{Xu}, \binits{M.H.}},
\oauthor{\bsnm{Soja}, \binits{B.}}:
Bridging astronomical, astrometric and geodetic scheduling for {VGOS}.
Journal of Geodesy
\textbf{97}(2)
(2023)
\doiurl{10.1007/s00190-023-01706-4}
\end{botherref}
\endbibitem

\bibitem[\protect\citeauthoryear{Haas et~al.}{2021}]{Haas2021}
\begin{barticle}
\bauthor{\bsnm{Haas}, \binits{R.}},
\bauthor{\bsnm{Varenius}, \binits{E.}},
\bauthor{\bsnm{Matsumoto}, \binits{S.}},
\bauthor{\bsnm{Schartner}, \binits{M.}}:
\batitle{{Observing UT1-UTC with VGOS}}.
\bjtitle{Earth, Planets and Space}
\bvolume{73}(\bissue{1}),
\bfpage{78}
(\byear{2021})
\doiurl{10.1186/s40623-021-01396-2}
\end{barticle}
\endbibitem

\bibitem[\protect\citeauthoryear{Luzum and Gambis}{2014}]{Luzum2014}
\begin{botherref}
\oauthor{\bsnm{Luzum}, \binits{B.}},
\oauthor{\bsnm{Gambis}, \binits{D.}}:
{Explanatory Supplement to IERS Bulleting A and Bulletin B/ C04}
(2014).
\url{https://hpiers.obspm.fr/iers/bul/bulb_new/bulletinb.pdf}
\end{botherref}
\endbibitem

\bibitem[\protect\citeauthoryear{Diamantidis and Haas}{2023}]{Diamantidis2023}
\begin{botherref}
\oauthor{\bsnm{Diamantidis}, \binits{P.-K.}},
\oauthor{\bsnm{Haas}, \binits{R.}}:
Evaluation of the first three years of vgos 24~h sessions using a kalman filter
  with c5++.
Earth, Planets and Space
\textbf{75}(1)
(2023)
\doiurl{10.1186/s40623-023-01872-x}
\end{botherref}
\endbibitem

\bibitem[\protect\citeauthoryear{{Glomsda} et~al.}{2023}]{Glomsda2023}
\begin{bchapter}
\bauthor{\bsnm{{Glomsda}}, \binits{M.}},
\bauthor{\bsnm{{Seitz}}, \binits{M.}},
\bauthor{\bsnm{{Angermann}}, \binits{D.}}:
\bctitle{{Comparison of Simultaneous VGOS and Legacy VLBI Sessions}}.
In: \beditor{\bsnm{{Armstrong}}, \binits{K.L.}},
\beditor{\bsnm{{Behrend}}, \binits{D.}},
\beditor{\bsnm{{Baver}}, \binits{K.D.}} (eds.)
\bbtitle{International VLBI Service for Geodesy and Astrometry 2022 General
  Meeting Proceedings},
pp. \bfpage{187}--\blpage{191}
(\byear{2023})
\end{bchapter}
\endbibitem

\bibitem[\protect\citeauthoryear{{Nilsson} et~al.}{2023}]{Nilsson2023}
\begin{bchapter}
\bauthor{\bsnm{{Nilsson}}, \binits{T.}},
\bauthor{\bsnm{{Haas}}, \binits{R.}},
\bauthor{\bsnm{{Varenius}}, \binits{E.}}:
\bctitle{{The Current and Future Performance of VGOS}}.
In: \beditor{\bsnm{{Armstrong}}, \binits{K.L.}},
\beditor{\bsnm{{Behrend}}, \binits{D.}},
\beditor{\bsnm{{Baver}}, \binits{K.D.}} (eds.)
\bbtitle{International VLBI Service for Geodesy and Astrometry 2022 General
  Meeting Proceedings},
pp. \bfpage{192}--\blpage{196}
(\byear{2023}).
\burl{https://ivscc.gsfc.nasa.gov/publications/gm2022/41_nilsson_etal.pdf}
\end{bchapter}
\endbibitem

\bibitem[\protect\citeauthoryear{Motlaghzadeh et~al.}{2022}]{Motlaghzadeh2022}
\begin{botherref}
\oauthor{\bsnm{Motlaghzadeh}, \binits{S.}},
\oauthor{\bsnm{Alizadeh}, \binits{M.M.}},
\oauthor{\bsnm{Cappallo}, \binits{R.}},
\oauthor{\bsnm{Heinkelmann}, \binits{R.}},
\oauthor{\bsnm{Schuh}, \binits{H.}}:
Deriving ionospheric total electron content by {VLBI} global observing system
  data analysis during the {CONT}17 campaign.
Radio Science
\textbf{57}(9)
(2022)
\doiurl{10.1029/2021rs007336}
\end{botherref}
\endbibitem

\bibitem[\protect\citeauthoryear{Haas et~al.}{2023}]{Haas2023}
\begin{barticle}
\bauthor{\bsnm{Haas}, \binits{R.}},
\bauthor{\bsnm{Diamantidis}, \binits{P.-K.}},
\bauthor{\bsnm{Elgered}, \binits{G.}},
\bauthor{\bsnm{Johansson}, \binits{J.}},
\bauthor{\bsnm{Nilsson}, \binits{T.}},
\bauthor{\bsnm{Ning}, \binits{T.}}:
\batitle{Assessment of parameters describing the signal delay in the neutral
  atmosphere derived from {VGOS} sessions}.
\bjtitle{EGU General Assembly 2023}
(\byear{2023})
\doiurl{10.5194/egusphere-egu23-16054}
\end{barticle}
\endbibitem

\bibitem[\protect\citeauthoryear{Nothnagel et~al.}{2017}]{Nothnagel2017}
\begin{barticle}
\bauthor{\bsnm{Nothnagel}, \binits{A.}},
\bauthor{\bsnm{Artz}, \binits{T.}},
\bauthor{\bsnm{Behrend}, \binits{D.}},
\bauthor{\bsnm{Malkin}, \binits{Z.}}:
\batitle{{International VLBI Service for Geodesy and Astrometry}}.
\bjtitle{Journal of Geodesy}
\bvolume{91}(\bissue{7}),
\bfpage{711}--\blpage{721}
(\byear{2017})
\doiurl{10.1007/s00190-016-0950-5}
\end{barticle}
\endbibitem

\bibitem[\protect\citeauthoryear{Schartner and B\"ohm}{2019}]{Schartner2019}
\begin{barticle}
\bauthor{\bsnm{Schartner}, \binits{M.}},
\bauthor{\bsnm{B\"ohm}, \binits{J.}}:
\batitle{{VieSched++: A New VLBI Scheduling Software for Geodesy and
  Astrometry}}.
\bjtitle{Publications of the Astronomical Society of the Pacific}
\bvolume{131}(\bissue{1002}),
\bfpage{084501}
(\byear{2019})
\doiurl{10.1088/1538-3873/ab1820}
\end{barticle}
\endbibitem

\bibitem[\protect\citeauthoryear{Schartner and B{\"o}hm}{2020}]{Schartner2020a}
\begin{barticle}
\bauthor{\bsnm{Schartner}, \binits{M.}},
\bauthor{\bsnm{B{\"o}hm}, \binits{J.}}:
\batitle{{Optimizing schedules for the VLBI global observing system}}.
\bjtitle{Journal of Geodesy}
\bvolume{94}(\bissue{1}),
\bfpage{12}
(\byear{2020})
\doiurl{10.1007/s00190-019-01340-z}
\end{barticle}
\endbibitem

\bibitem[\protect\citeauthoryear{Schartner et~al.}{2021}]{Schartner2021b}
\begin{barticle}
\bauthor{\bsnm{Schartner}, \binits{M.}},
\bauthor{\bsnm{Pl{\"o}tz}, \binits{C.}},
\bauthor{\bsnm{Soja}, \binits{B.}}:
\batitle{{Automated VLBI scheduling using AI-based parameter optimization}}.
\bjtitle{Journal of Geodesy}
\bvolume{95}(\bissue{5}),
\bfpage{58}
(\byear{2021})
\doiurl{10.1007/s00190-021-01512-w}
\end{barticle}
\endbibitem

\bibitem[\protect\citeauthoryear{B\"{o}hm et~al.}{2018}]{Boehm2018}
\begin{barticle}
\bauthor{\bsnm{B\"{o}hm}, \binits{J.}},
\bauthor{\bsnm{B\"{o}hm}, \binits{S.}},
\bauthor{\bsnm{Boisits}, \binits{J.}},
\bauthor{\bsnm{Girdiuk}, \binits{A.}},
\bauthor{\bsnm{Gruber}, \binits{J.}},
\bauthor{\bsnm{Hellerschmied}, \binits{A.}},
\bauthor{\bsnm{Kr{\'{a}}sn{\'{a}}}, \binits{H.}},
\bauthor{\bsnm{Landskron}, \binits{D.}},
\bauthor{\bsnm{Madzak}, \binits{M.}},
\bauthor{\bsnm{Mayer}, \binits{D.}},
\bauthor{\bsnm{McCallum}, \binits{J.}},
\bauthor{\bsnm{McCallum}, \binits{L.}},
\bauthor{\bsnm{Schartner}, \binits{M.}},
\bauthor{\bsnm{Teke}, \binits{K.}}:
\batitle{{Vienna VLBI and Satellite Software (VieVS) for Geodesy and
  Astrometry}}.
\bjtitle{Publications of the Astronomical Society of the Pacific}
\bvolume{130}(\bissue{986}),
\bfpage{044503}
(\byear{2018})
\doiurl{10.1088/1538-3873/aaa22b}
\end{barticle}
\endbibitem

\bibitem[\protect\citeauthoryear{{Bolotin} et~al.}{2014}]{Bolotin2014}
\begin{bchapter}
\bauthor{\bsnm{{Bolotin}}, \binits{S.}},
\bauthor{\bsnm{{Baver}}, \binits{K.}},
\bauthor{\bsnm{Gipson}, \binits{J.}},
\bauthor{\bsnm{D.}, \binits{G.}},
\bauthor{\bsnm{D.}, \binits{M.}}:
\bctitle{{The VLBI Data Analysis Software nuSolve: Development Progress and
  Plans for the Future}}.
In: \beditor{\bsnm{{Behrend}}, \binits{D.}},
\beditor{\bsnm{{Baver}}, \binits{K.D.}},
\beditor{\bsnm{{Armstrong}}, \binits{K.L.}} (eds.)
\bbtitle{International VLBI Service for Geodesy and Astrometry 2014 General
  Meeting Proceedings: ``VGOS: The New VLBI Network'', Eds. Dirk Behrend, Karen
  D. Baver, Kyla L. Armstrong, Science Press, Beijing, China, ISBN
  978-7-03-042974-2, 2014, P. 267-271},
pp. \bfpage{253}--\blpage{257}
(\byear{2014})
\end{bchapter}
\endbibitem

\bibitem[\protect\citeauthoryear{Kr\'asn\'a et~al.}{2023}]{Krasna2023a}
\begin{botherref}
\oauthor{\bsnm{Kr\'asn\'a}, \binits{H.}},
\oauthor{\bsnm{Baldreich}, \binits{L.}},
\oauthor{\bsnm{B\"ohm}, \binits{J.}},
\oauthor{\bsnm{B\"ohm}, \binits{S.}},
\oauthor{\bsnm{Gruber}, \binits{J.}},
\oauthor{\bsnm{Hellerschmied}, \binits{A.}},
\oauthor{\bsnm{Jaron}, \binits{F.}},
\oauthor{\bsnm{Kern}, \binits{L.}},
\oauthor{\bsnm{Mayer}, \binits{D.}},
\oauthor{\bsnm{Nothnagel}, \binits{A.}},
\oauthor{\bsnm{Panzenb\"ock}, \binits{O.}},
\oauthor{\bsnm{Wolf}, \binits{H.}}:
{VLBI} celestial and terrestrial reference frames {VIE2022b}.
Astronomy \& Astrophysics
\textbf{accepted}
(2023)
\doiurl{10.48550/arXiv.2211.07338}
\end{botherref}
\endbibitem

\bibitem[\protect\citeauthoryear{Kr\'asn\'a}{2023}]{Krasna2023}
\begin{botherref}
\oauthor{\bsnm{Kr\'asn\'a}, \binits{H.}}:
{Earth orientation parameters (EOP) from VIE2022 VLBI solution}.
TU Wien.
(2.0.0) [Data set]
(2023).
\doiurl{10.48436/0gmbv-arv60} .
\url{https://doi.org/10.48436/0gmbv-arv60}
\end{botherref}
\endbibitem

\end{thebibliography}

\end{document}